\documentclass{article}
\usepackage{amssymb}
\usepackage{xcolor}
\usepackage{amsmath}
\usepackage{makecell}
\usepackage{amsthm,subfigure,graphics,epsfig}
\usepackage{epstopdf}
\usepackage{setspace}
\newcommand{\sign}{\text{sign}}
\usepackage{amsfonts,amstext,amsmath,amssymb,amsthm}
\usepackage{mathtools}
\usepackage[ruled,vlined]{algorithm2e}
\def\spacingset#1{\renewcommand{\baselinestretch}%
{#1}\small\normalsize} \spacingset{1}
\usepackage{arxiv}

\usepackage[utf8]{inputenc} 
\usepackage[T1]{fontenc}    
\usepackage{hyperref}       
\usepackage{url}            
\usepackage{booktabs}       
\usepackage{amsfonts}       
\usepackage{nicefrac}       
\usepackage{microtype}      
\usepackage{lipsum}		
\usepackage{graphicx}
\usepackage{natbib}
\usepackage{doi}

\title{Deep Learning for Efficient GWAS Feature Selection}

\author{ {\hspace{1mm}Kexuan Li} \\
	Global Biometrics and Data Sciences\\
	Bristol Myers Squibb\\
	\texttt{kexuan.li.77@gmail.com} \\
}

\renewcommand{\shorttitle}{\textit{arXiv} Template}

\begin{document}
\maketitle

\begin{abstract}
Genome-Wide Association Studies (GWAS) face unique challenges in the era of big genomics data, particularly when dealing with ultra-high-dimensional datasets where the number of genetic features significantly exceeds the available samples. This paper introduces an extension to the feature selection methodology proposed by Mirzaei et al. (2020), specifically tailored to tackle the intricacies associated with ultra-high-dimensional GWAS data. Our extended approach enhances the original method by introducing a Frobenius norm penalty into the student network, augmenting its capacity to adapt to scenarios characterized by a multitude of features and limited samples. Operating seamlessly in both supervised and unsupervised settings, our method employs two key neural networks. The first leverages an autoencoder or supervised autoencoder for dimension reduction, extracting salient features from the ultra-high-dimensional genomic data. The second network, a regularized feed-forward model with a single hidden layer, is designed for precise feature selection. The introduction of the Frobenius norm penalty in the student network significantly boosts the method's resilience to the challenges posed by ultra-high-dimensional GWAS datasets. Experimental results showcase the efficacy of our approach in feature selection for GWAS data. The method not only handles the inherent complexities of ultra-high-dimensional settings but also demonstrates superior adaptability to the nuanced structures present in genomics data. The flexibility and versatility of our proposed methodology are underscored by its successful performance across a spectrum of experiments.
\end{abstract}


\spacingset{1.25}

\section{Introduction}
Feature selection stands as a critical cornerstone in numerous biological studies, a process pivotal in unraveling the complexities of data-intensive domains such as genome-wide association studies (GWAS), microarray analysis, and mass spectra analysis. The omnipresent challenge lies in datasets characterized by an inherent high-dimensional nature, coupled with a paucity of observations. In the realm of GWAS, a paradigmatic exploration where the identification of genes and the delineation of associations between single-nucleotide polymorphisms (SNPs) and human diseases take center stage, the landscape is riddled with obstacles. The GWAS datasets often manifest a disconcerting dichotomy — an expansive number of SNPs (e.g., $p\geq10^5$) juxtaposed against a relatively diminutive sample size (e.g., $n\leq10^3$). Navigating through this ultra-high-dimensional space and extracting a representative set of SNPs emerges as a persistent and formidable challenge in the quest for deciphering genetic underpinnings.

Numerous methodologies have been proposed to address the challenge of feature selection in GWAS data. To analysize GWAS data, the Cochran-Armitage trend test \citep{cochran1954some, armitage1955tests} has become a standard procedure for association testing in large-scale genome-wide association studies. However, various more complicated models had been developed to analize GWAS data as well. For instance, \cite{SIS} introduced the Marginal Sure Independence Screening procedure (SIS), specifically designed for ultrahigh-dimensional linear models, relying on Pearson correlations. Subsequent efforts in feature screening have yielded diverse procedures tailored to various models and successfully applied to GWAS data \citep{cui2015model}. Another prominent avenue in GWAS feature selection leverages the power of Lasso. In studies such as \citep{LassoWu} and \citep{ayers2010snp}, penalized logistic regression underpinned by Lasso has been employed to unravel associations within GWAS data. Exploring Lasso coefficients, \cite{arbet2017resampling} investigated alternative, swift, and potent methods, highlighting the efficacy of permutation selection and analytic selection as alternatives to standard univariate analysis in GWAS data. Addressing the intricacies of joint multiple-SNP regression models, \cite{pLasso} proposed a permutation-assisted tuning procedure within the Lasso framework to discern phenotype-associated SNPs. Beyond these Lasso-based models, a myriad of feature selection methods tailored for GWAS data exists. For example, \cite{de2014snps} proposed a methodology to simultaneously select the most relevant SNPs markers
for the characterization of any measurable phenotype described by a continuous variable using Support Vector
Regression with Pearson Universal kernel. \cite{li2018predicting} harnessed incremental feature selection to unearth novel gene expression patterns in brain tissues associated with early wake-up, drawing insights from GWAS data. \cite{cueto2019comparative} provided a comprehensive comparative study on various machine learning based feasure selection methods on or colorectal cancer. Meanwhile, \cite{chu2020feature} proposed a two-step gene-detection procedure embedded in generalized varying coefficient mixed-effects models. For a more exhaustive exploration of these approaches, please refer to the comprehensive review by \citep{tadist2019feature, pudjihartono2022review}. Despite their utility, classical methods encounter challenges in large biological datasets, including:

\begin{itemize}
\item \textbf{Feature Dependencies and Nonlinear Structures}:\\
Traditional feature selection methods, broadly categorized as filter, wrapper, and embedded methods, exhibit limitations that become pronounced in the intricate landscape of GWAS datasets. For instance, filter methods often make assumptions about parametric model forms (e.g., linear models, lasso-based models, logistic regression models) and tend to overlook intricate interactions between features or nonlinear structures \citep{LassoWu, pLasso}.

\item \textbf{Lack of Flexibility}:\\
Moreover, the rigidity of many algorithms poses a substantial roadblock. While the spotlight often shines on supervised feature selection, the realm of unsupervised feature selection is equally pivotal in biology. Consider, for example, clustering analysis, where the objective is to unearth new phenotypes by selecting genes devoid of prior phenotype knowledge \citep{feasure_selection_survey_unsurpervised_1, feasure_selection_survey_unsurpervised_2}.

\item \textbf{Dealing with Unbalanced Data and Reconstruction Challenges}:\\
The nuances of unbalanced data, an omnipresent challenge in large biological datasets, introduce complexities \citep{Imbalanced_1}. Additionally, the classical methods grapple with the intricate task of reconstruction and imputation, adding layers of intricacy to the feature selection process.
\end{itemize}

To address these challenges, we build upon the approach introduced by \citep{mirzaei2020deep}, originally designed for feature selection but not explicitly tailored for ultra-high-dimensional data. Recognizing the unique demands of ultra-high-dimensional settings, particularly prevalent in GWAS data, we extend their method to enhance its applicability to datasets with a large number of features. Specifically, we introduce a Frobenius norm penalty into the student network, adapting the approach to better navigate the complexities associated with ultra-high-dimensional and small-sample scenarios. Our extended method maintains its flexibility, proving effective in both supervised and unsupervised scenarios, and excelling at uncovering intricate nonlinear structures and interactions within the data. The architecture of our method comprises two neural networks: the first dedicated to dimension reduction through an autoencoder or supervised autoencoder, and the second utilizing a regularized feed-forward network with only one hidden layer. This extension refines the original approach, making it well-suited for the challenges posed by ultra-high-dimensional datasets encountered in GWAS analyses.

The remainder of the paper is structured as follows: Section \ref{Sec2} delineates the problem of interest and offers a comprehensive literature review. Section \ref{SecMethod} provides an in-depth exposition of the proposed method. In Section \ref{Sec4}, we apply the proposed method and conduct comparisons with alternative approaches across various experiments. Finally, Section \ref{Secconclusion} delves into a discussion of the proposed method.

\section{Problem of Interest and Literature Review} \label{Sec2}
\subsection{Problem Formulation}
Let's delve into the intricacies of both supervised and unsupervised feature selection within the context of GWAS. Imagine a set of observations $\boldsymbol{x}_i \in \mathbb{R}^p, i=1, \ldots, n$, representing a sample size $n$ with $p$ features, assumed to be independent and identically distributed (i.i.d.) from a distribution $p(\boldsymbol{x})$. In the realm of unsupervised feature selection specific to GWAS, the goal is to discern a subset $\mathcal{S}  \subseteq \{1, 2, \ldots, p\}$ comprising the most discriminative and informative features. This subset, with $|\mathcal{S}| = k \leqslant p$, is accompanied by a reconstruction function $f: \mathbb{R}^k \rightarrow \mathbb{R}^p$. The critical aspect here is the mapping from a low-dimensional feature space $\mathbb{R}^k$ back to the original feature space $\mathbb{R}^p$. The aim is to minimize the expected loss between $f(\boldsymbol{x}^{(\mathcal{S})})$ and the original input $\boldsymbol{x}$, where $\boldsymbol{x}^{(\mathcal{S})} = (x_{s_1}, \ldots, x_{s_k})^{\top} \in \mathbb{R}^k$ and $s_i \in \mathcal{S}$ represents the low-dimensional $k$ features. This process is pivotal in GWAS, where selecting relevant genetic features and understanding their intricate relationships contribute significantly to uncovering associations with complex traits and diseases. Moving to supervised feature selection, the complexity increases with the availability of both the sample design matrix $\boldsymbol{X}=(\boldsymbol{x}_1, \ldots,\boldsymbol{x}_n)^{T} \in \mathbb{R}^{n \times p}$ and the label vector $\boldsymbol{y} = (y_1,\ldots,y_n)^T \in \mathbb{R}^n$. Here, $y_i$ can be continuous or categorical, and the assumption is an unknown true relationship between a subset of features and $y$, expressed as $y = f(\boldsymbol{x}^{(\mathcal{S})})$. This relationship involves $\mathcal{S} \subseteq \{1, 2, \ldots, p\}$ with $|\mathcal{S}| = k \leqslant p$ and $\boldsymbol{x}^{(\mathcal{S})} \in \mathbb{R}^k$. Traditional linear assumptions, such as $y_i = f(\boldsymbol{x}_i^{(\mathcal{S})}) = \beta_0 + \sum_{j=1}^k\beta_j x_{i, s_j} + \epsilon_i$ for continuous responses or logistic regression $\log \frac{Pr(y_i = 1)}{Pr(y_i = 0)} = \beta_0 + \sum_{j=1}^k\beta_j x_{i, s_j}$ for categorical responses, are common in GWAS. However, the limitations of such linear models prompt the exploration of more expressive choices like neural networks. In the realm of GWAS, this adaptive approach is crucial for capturing the intricate genetic architecture underlying complex phenotypes.
\subsection{Related Works}
In recent years, the remarkable success of DNNs has reverberated across a myriad of domains, underscoring their versatility and potency. These domains span from the visually immersive realms of computer vision \citep{he2016deep} to the intricate intricacies of deciphering language in natural language processing \citep{bahdanau2014neural}. DNNs have left an indelible mark on recommendation systems \citep{zhang2019deep}, offering personalized suggestions with unprecedented accuracy, and have even delved into the realms of drug discovery \citep{jimenez2020drug}, spatial data analytics \citep{li2023semiparametric}, computational biology \citep{angermueller2016deep}, and the nuanced dynamics of complex systems \citep{li2021calibrating}. Amidst this expansive landscape, the application of DNNs to feature selection has emerged as a fascinating area of exploration, garnering significant attention from researchers. The integration of DNNs with sparse group lasso to tackle problems of heterogeneous feature representations was a pivotal exploration by \citep{zhao2015heterogeneous}. Meanwhile, \cite{li2016deep} introduced Deep Feature Selection (DFS), a novel approach employing regularization techniques to rank feature importance, contributing to the nuanced understanding of feature relevance in complex datasets. In the pursuit of effective feature selection, \cite{liu2017deep} proposed a strategy known as deep neural pursuit (DNP). This method strategically selects relevant features by leveraging the averaging out of gradients, employing multiple dropouts to lower variance and enhance robustness. Building upon the foundation of the knockoffs framework, \cite{lu2018deeppink} incorporated this methodology into DNNs, enabling feature selection while maintaining a controlled error rate—an essential consideration in applications where precision is paramount. In the realm of high-dimensional nonlinear variable selection, \cite{chen2021nonlinear} established a comprehensive framework utilizing DNNs. They demonstrated the method's selection consistency under the condition of a generalized stable restricted Hessian in the objective function. This breakthrough contributes significantly to the understanding and application of DNNs in scenarios characterized by complex, nonlinear relationships between variables. Adding to this rich tapestry of DNN applications, \cite{gui2019afs} introduced an attention-based mechanism for supervised feature selection. This mechanism harnesses the power of attention in neural networks to dynamically emphasize and de-emphasize features, providing a nuanced approach to selecting relevant variables in a supervised context. \cite{abid2019concrete,singh2023fsnet}, who ingeniously harnessed the Gumble-Softmax trick \citep{jang2016categorical}. They introduced a concrete selector layer into the architecture, allowing gradients to seamlessly pass through the network during the feature selection process. Further enriching the landscape, \cite{lemhadri2021lassonet} presented LassoNet, a feature selection network that introduced a residual layer between the input and output layers. Notably, this architecture imposed penalties on parameters within the residual layer while ensuring that the norm of parameters in the first layer remained less than the corresponding norm in the residual layer.  Building upon the foundations laid by LassoNet, \cite{li2022variable} extended its applicability to censored data. Additionally, \cite{li2023deep} introduced a comprehensive approach that integrates deep learning and feature screening, yielding a framework capable of achieving both supervised and unsupervised feature selection. This hybrid methodology capitalizes on the strengths of both paradigms.
Nevertheless, directly applying standard DNNs to GWAS data presents challenges. On one hand, interpretability is paramount in biological data, and many deep learning algorithms operate as "black-boxes," lacking inherent interpretability. On the other hand, the efficacy of most deep learning algorithms hinges on abundant training data, a luxury often unattainable in biological and medical datasets. These domains frequently contend with smaller sample sizes in comparison to the data dimension $(p >> n)$, posing a distinct set of challenges.

\section{Methods}\label{SecMethod}
We present a two-stage deep neural network designed to effectively capture the intricate structure of genomic data. Our first-stage neural network, chosen for its complexity, employs a supervised autoencoder to ensure an expressive model capable of extracting the complex manifold inherent in the data. This complexity is essential for capturing nonlinear structures within the features, a task unattainable by simpler models like linear ones. In cases where a response variable is unavailable (unsupervised scenarios), our approach seamlessly transforms the supervised autoencoder into a standard autoencoder. The primary goal of the first stage is dimensionality reduction and feature extraction. Following the training of the first stage, we transform the high-dimensional input into a low-dimensional feature space, which serves as the output for the second stage. The second stage employs a single fully-connected layer with sparsity regularization to reproduce the output from the first stage. Utilizing sparsity regularization on the weight matrix facilitates feature selection by emphasizing the highest feature scores. Comprehensive details for each stage are elaborated in the ensuing sections.

\subsection{The First Stage: Dimension Reduction}
In the initial stage, a supervised autoencoder is employed to acquire a sophisticated representation of the input data. In scenarios where labels are unavailable (unsupervised situations), a conventional autoencoder is seamlessly substituted. Autoencoders, a distinct class of feed-forward neural networks, specialize in dimension reduction. Notably, many traditional dimension reduction techniques can be considered specific instances of autoencoders. For instance, Principal Component Analysis (PCA) can be conceptualized as an autoencoder when the loss function corresponds to mean square loss without an activation function.

A standard (unsupervised) autoencoder consists of two parts, the encoder and the decoder. Suppose the input space and output space is $\mathcal{X}$, the hidden layer space is $\mathcal{F}$. The goal is to find two maps $\Phi:\mathcal{X}\rightarrow \mathcal{F}; \Psi:\mathcal{F}\rightarrow \mathcal{X} $ which minimize the loss $\mathcal{L}_r(\Theta|\boldsymbol{X}) = \frac{1}{n}\sum_{i=1}^n||\boldsymbol{x}_i - \Psi(\Phi(\boldsymbol{x}_i))||_2^2 $, where $\mathcal{L}_r(\cdot)$ is the reconstruction loss function and $\Theta = [\Theta_\Phi, \Theta_\Psi]$ are the model parameters. Here, $\Phi$ is called the encoder, and $\Psi$ is called the decoder. To better learn the non-linear structure of features, people always assume $\Phi$ and $\Psi$ are neural networks. For example, if there is only one layer in both encoder and decoder with mean square loss, then $\Phi(\boldsymbol{x})=\sigma(\boldsymbol{W}\boldsymbol{x}+b)\in \mathcal{F}, \Psi(\Phi(\boldsymbol{x}))=\sigma'(\boldsymbol{W}'\Phi(\boldsymbol{x})+b')$, and
\begin{equation} \label{Eq Reconstruction loss}
\mathcal{L}_r(\Theta|\boldsymbol{X}) = \frac{1}{n}\sum_{i=1}^n||\boldsymbol{x}_i - \Psi(\Phi(\boldsymbol{x}_i))||_2^2 = \frac{1}{n}\sum_{i=1}^n||\boldsymbol{x}_i - \sigma'(\boldsymbol{W}'\sigma(\boldsymbol{W}\boldsymbol{x}_i+b)+b')||_2^2,
\end{equation}
where $\sigma, \sigma'$ are nonlinear active functions, $\boldsymbol{W}, \boldsymbol{W}'$ are weight matrices, $b, b'$ are bias vectors, and $||\cdot||_2$ is the $l_2$ norm. The standard autoencoder can be extended to many other forms, such as sparse autoencoder (SAE), denoising autoencoder (DAE), variational autoencoder (VAE).

In this paper, since we focus on the supervised feature selection, we will use a supervised autoencoder instead of the standard (unsupervised) autoencoder. In supervised autoencoder, we add an additional loss on the hidden layer, for example, mean square loss for continuous response or cross-entropy loss for categorical response. Let $\mathcal{L}_s(\cdot)$ be the supervised loss on the hidden layer and $\mathcal{L}_r(\cdot)$ be the reconstruction loss as in Equation (\ref{Eq Reconstruction loss}). The loss in the supervised autoencoder with continuous response is:
\begin{equation} \label{Eq Continuous loss}
\mathcal{L}(\Theta_1|\boldsymbol{X}, \boldsymbol{y}) = \mathcal{L}_s(\Theta_\Phi, \Theta_\Upsilon|\boldsymbol{X}, \boldsymbol{y}) + \mathcal{L}_r(\Theta_\Phi, \Theta_\Psi|\boldsymbol{X})
                     = \frac{1}{n}\sum_{i=1}^n\left(||y_i -  \Upsilon(\Phi(\boldsymbol{x}_i))||_2^2 + \lambda||\boldsymbol{x}_i - \Psi(\Phi(\boldsymbol{x}_i))||_2^2 \right),
\end{equation}
where $\Theta_1 = [\Theta_\Phi, \Theta_\Psi, \Theta_\Upsilon]$ is the model parameters, $\Phi$ is the encoder, $\Psi$ is the decoder, $\Upsilon$ is the regressor, and $\lambda$ is the regularization parameter for the reconstruction loss. The corresponding loss in the supervised autoencoder with categorical response is:
\begin{equation} \label{Eq Coategorical loss}
\begin{split}
\mathcal{L}(\Theta_1|\boldsymbol{X}, \boldsymbol{y}) &= \mathcal{L}_s(\Theta_\Phi, \Theta_\Upsilon|\boldsymbol{X}, \boldsymbol{y}) + \mathcal{L}_r(\Theta_\Phi, \Theta_\Psi|\boldsymbol{X})\\
                     &= \frac{1}{n}\sum_{i=1}^n\left(-\log\left(\frac{\exp(\Upsilon(\boldsymbol{x}_i)_{y_i})}{\sum_{c=1}^C\exp(\Upsilon(\boldsymbol{x}_i)_c)}\right) + \lambda||\boldsymbol{x}_i - \Psi(\Phi(\boldsymbol{x}_i))||_2^2 \right),
\end{split}
\end{equation}
where $\Theta_1, \Phi$, $\Psi$ and $\lambda$ are the same as Equation (\ref{Eq Continuous loss}), $C$ is the number of classes, and $\Upsilon$ is a classifier on the hidden layer with softmax output. The training process in the first stage is an optimization problem to minimize the loss function $\mathcal{L}(\Theta_1 | \boldsymbol{X}, \boldsymbol{y})$.

Once the model is trained, we can extract the features by mapping the original input from $\mathcal{X}$
to the low dimension hidden space $\mathcal{F}$. Without loss of generality, we assume $\mathcal{X} = \mathbb{R}^p, \mathcal{F}=\mathbb{R}^h$ where $h  \ll p$. Define normalized encoded input
\[
\boldsymbol{x}_{\text{encode}} = \frac{\Phi(\boldsymbol{x}) - \min\Phi(\boldsymbol{x})}{\max\Phi(\boldsymbol{x}) - \min \Phi(\boldsymbol{x})} \in \mathcal{F},
\]
which generates the abstract features from the original high-dimensional data and will be used in the second stage.

\subsection{The Second Stage: Feature Selection}
\begin{figure}[h]
\label{Structure}
\centering
\includegraphics[scale=0.72]{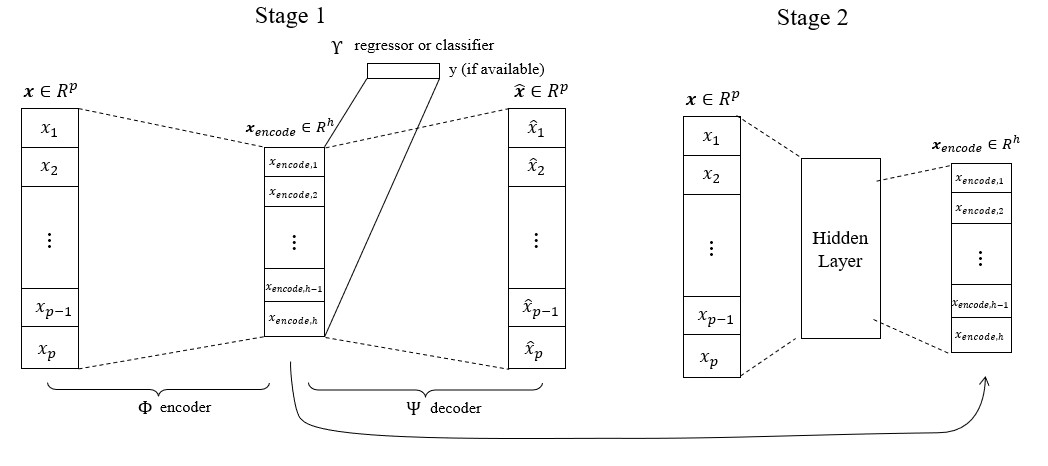}
\caption{The architecture of the proposed DNN consists of an autoencoder based dimension reduction
method for feature extraction (on the left) and a feature selection network (on the right).
$\boldsymbol{x}_{\textrm{encode}} \in \mathbb{R}^h$ is the low-dimensional representation of the original input that captures most of the information of the data in $\mathbb{R}^p$, $h \ll p$, and is used to compute the strength of dependence with each feature. The second stage is a single-layer neural network with a row-sparse regularization and a weight decay term on the weight matrix to mimic the $\boldsymbol{x}_{\textrm{encode}}$ from the first stage.}
\label{Flowchart}
\end{figure}

In the second stage, we train a single-layer neural network with a row-sparse regularization and a weight decay term on the weight matrix to mimic the $\boldsymbol{x}_\text{encode}$ from the first stage. The reason why we use a simple neural network is because we want to make sure the gradient can be easily back-propagated to the first layer such that the most important features can be successfully selected.  To be more specific, the neural network in the second stage is defined as $\hat{\boldsymbol{x}} = \boldsymbol{W_2}\left(\sigma(\boldsymbol{W}_1\boldsymbol{x}+b_1)\right)+b_2$, where $\boldsymbol{W}_1, \boldsymbol{W}_2, b_1, b_2$ are the weight matrices and biases of the first layer and the output layer and $\sigma(\cdot)$ is the activation function. The loss function we want to optimize is
\begin{equation}
\mathcal{L}(\Theta_2|\boldsymbol{X}, \boldsymbol{X}_\text{encode})
= \frac{1}{n}\sum_{i=1}^n ||\hat{\boldsymbol{x}}-\boldsymbol{x}_\text{encode}||_2^2 +\alpha||\boldsymbol{W}_1||_{2,1} +\frac{\beta}{2}\sum_{i=1}^2||\boldsymbol{W}_i||_F^2,
\end{equation}
where $\Theta_2 = [\boldsymbol{W}_1, \boldsymbol{W}_2, b_1, b_2]$, $\alpha, \beta$ are penalty parameters,
\[
||\boldsymbol{W}_{m\times n}||_{2,1} = \sum_{j=1}^n\left(\sum_{i=1}^m|w_{ij}|^2\right)^\frac{1}{2} ,\,\, ||\boldsymbol{W}_{m\times n}||_F = \sqrt{\sum_{i=1}^m\sum_{j=1}^n|W_{ij}|^2}=\sqrt{\mbox{trace}(W^TW)}
\]
are the $L_{2,1}$ norm and the Frobenius norm, respectively. We add a row-sparse regularization term ($||\cdot||_{2,1}$) in the first layer for feature selection and weight decay term (Frobenius norm) to avoid overfitting and help convergence. Once the second stage neural network is trained, we can define the feature scores as
\[
s={\mbox{diag}(W_1W_1^T)},
\]
where $s$ is a $p$ dimensional vector of feature importance. The larger the feature importance is, the more significant role that feature plays. The idea of adding a row-sparse regularization to
hidden layers in feature selection has also be investigated in literature \citep{scardapane2017group, han2018autoencoder, feng2018graph}. The algorithm and the architecture of the method are summarized in Algorithm \ref{Algo} and Figure \ref{Structure}.
\subsection{IMPLEMENTATION}

The minimization of loss functions typically employs stochastic gradient descent (SGD), updating model parameters based on gradients computed from small, randomly drawn batches (usually 32 to 512). These batches significantly reduce the computational cost of gradient calculations. In the context of large-scale GWAS data, we utilize a scalable algorithm to optimize the loss function $\mathcal{L}$ through stochastic gradient descent. Our method encompasses crucial hyperparameters, including the number of layers, neurons per layer, dropout probability, learning rate, regularization parameters, and more. Configuring these hyperparameters accurately is vital for achieving optimal performance. However, determining suitable values can be challenging without domain expertise. Common strategies, such as grid search, random search (\citep{RandomSearch}), and Bayesian optimization (\citep{Bayesian_Opt}), are often employed in practice. In our approach, we adopt a naive search, evaluating the loss for predefined parameter candidates based on a validation set.

\begin{algorithm}
\label{Algo}
\KwIn{input design matrix $\boldsymbol{X}\in\mathbb{R}^{n\times p}$, labels $y\in\{1,...,C\}$ (if available), encoder network $\Phi$, decoder network network $\Psi$, regressor or classifier $\Upsilon$ (if available), learning rate, penalty term $\lambda$, regularization terms $\alpha, \beta$, and number of epochs for each stage $E_1, E_2$.}
\KwOut{feature scores}
\textbf{Train the First Stage Neural Network:}\\
Initialize $\Theta_1 = [\Theta_\Phi, \Theta_\Psi, \Theta_\Upsilon].$\\
\For {$e \in \{1,...,E_1\}$}{
$\hat{\boldsymbol{x}}_\text{encode} = \Phi(x)$\\
$\hat{y} = \Upsilon(\hat{\boldsymbol{x}}_\text{encode})$ (if labels are available)\\
$\hat{\boldsymbol{x}} = \Psi(\hat{\boldsymbol{x}}_\text{encode})$\\
\If {response y is not available:}  {$\mathcal{L} = ||\hat{\boldsymbol{x}}||_2^2$}
\If {response y is continuous:}  {$\mathcal{L} = \frac{1}{n}\sum_{i=1}^n\left(||y_i -  \Upsilon(\Phi(\boldsymbol{x}_i))||_2^2 + \lambda||\boldsymbol{x}_i - \Psi(\Phi(\boldsymbol{x}_i))||_2^2 \right)$}
\If {response y is categorical:}  {$\mathcal{L} =  \frac{1}{n}\sum_{i=1}^n\left(-\log\left(\frac{\exp(\Upsilon(\boldsymbol{x}_i)_{y_i})}{\sum_{c=1}^C\exp(\Upsilon(\boldsymbol{x}_i)_c)}\right) + \lambda||\boldsymbol{x}_i - \Psi(\Phi(\boldsymbol{x}_i))||_2^2 \right)$}
optimize the loss function using RMSprop.
}
Finish Training the First Stage Neural Network. \\
Map input into to hidden space: $\boldsymbol{x}_\text{encode} = \Phi(x)$.
\caption{Two Stage Deep Feature Selection Network}
\textbf{Train the Second Stage Neural Network:}\\
Initialize $\Theta_2 = [\boldsymbol{W}_1, \boldsymbol{W}_2, b_1, b_2]$\\
\For {$e \in \{1,...,E_2\}$}{
$\hat{\boldsymbol{x}} = \boldsymbol{W}_2\left(\sigma(\boldsymbol{W}_1\boldsymbol{x}+b_1)\right)+b_2$\\
$\mathcal{L} = ||\hat{\boldsymbol{x}}-\boldsymbol{x}_\text{encode}||_2^2 +\alpha||\boldsymbol{W}_1||_{2,1} +\frac{\beta}{2}\sum_{i=1}^2||\boldsymbol{W}_i||_F^2$
optimize the loss function using RMSprop.
}
Finish Training the Second Stage Neural Network. \\
\KwRet $s={\mbox{diag}(W_1W_1^T)}$.
\end{algorithm}

\enlargethispage{6pt}

\section{Experiments}\label{Sec4}
In this section, we apply the proposed method to several experiments, with a particular emphasis on scenarios where the response variable is binary, representing dichotomous phenotypes. It's important to note that our approach is not limited solely to GWAS data and can be effectively employed in broader contexts, including both supervised feature selection and unsupervised feature selection.

\subsection{Experiment 1}
We follow the design of a previous GWAS feature selection paper \citep{pLasso}. Suppose the number of observations is $n=200, 500, 1000$ with dimension $p=1000, 2000, 5000$. Let $y \in \{0, 1\}$ be the binary response and $\boldsymbol{x} = (x_1, \ldots, x_p)^T, x_i \in \{0, 1, 2\}, i=1, \ldots, p$ be the SNPs. The data generation process is described as below. We first generated $n$ independent $p$-dimensional random variable $\boldsymbol{Z}_i=(Z_{i1}, \ldots, Z_{ip})^T, i=1, \ldots, n$, following a multivariate Gaussian distribution $\mathcal{N}(0, \boldsymbol{\Sigma})$, where $\boldsymbol{\Sigma}$ is a $p \times p$ covariance matrix to capture the correlation between SNPs. The design of $\boldsymbol{\Sigma}$ is the same as \citep{pLasso}. That is
\begin{equation*}
    \begin{aligned}
    \boldsymbol{\Sigma}_{ij} =
            \left\{
             \begin{array}{ll}
              1   & i=j,\\
              \rho   & i\neq j, i,j \leq p/20, \\
              0   & \textit{eotherwise},
             \end{array}
            \right.
    \end{aligned}
\end{equation*}
and $\rho =0, 0.4, 0.8$. This design allows closer SNPs having a stronger correlation. Next, we randomly generate $p$ minor allele frequencies (MAFs) $m_1, \ldots, m_p$ from the uniform distribution $Uniform(0.05, 0.5)$ to represent the strength of heritability. For the $i$-th observation $(y_i, \boldsymbol{x}_i), i=1, \ldots ,n$, the SNPs are generated be the following rule:
\begin{equation*}
    \begin{aligned}
    x_{ij} =
            \left\{
             \begin{array}{ll}
              0   &Z_{ij} \leq c_1,\\
              1   & c_1 < Z_{i,j} < c_2, \\
              2   &Z_{i,j} \geq c_2,
             \end{array}
            \right.
    \end{aligned}
\end{equation*}
where $c_1, c_2$ are the $(1-m_j)^2$-quantile and $(1-m_j^2)$-quantile of $\{Z_{1j}, \ldots, Z_{nj}\}, j=1, \ldots p$, respectively. We standardize each SNP to have a mean
of zero and a variance of one. It is worth mentioning that this data generation process is the same as \citep{pLasso}.

We further define a set $\mathcal{J} = \{j_1, \ldots, j_{10}\}$, where $j_k, k=1,\ldots,10$, are randomly sampled from $\{1, \ldots, p\}$ without replacement. The phenotype $y$ is generated according to the dichotomous phenotype model
\begin{align*}
\log\frac{\pi}{1-\pi} &= -3 + \beta_1x_{j_1} + \beta_2\sin(x_{j_2} + x_{j_1}) + \beta_3\log(x_{j_3}^2+1) + \beta_4x_{j_4}^2\\
 & + \beta_5\sign(x_{j_5}-1) + \beta_6\max(x_{j_6} + x_{j_8}, 2) + \beta_7x_{j_7}\sign(x_{j_7}-1)\\
 & + \beta_8\sqrt{|x_{j_8}-1|} + \beta_9\cos(x_{j_9}) + \beta_{10}\tanh(x_{j_{10}})
\end{align*}
where $\beta_j \sim Uniform(1, 2)$, for $j=1, \ldots, 10$, and $y\sim Binomial(1; \pi)$. In other words, there are $10$ out of $p$ active SNPs associated with phenotype.

We compare our method with other four methods:  Armitage trend test (ATT), iterative SIS \citep{SIS}, Lasso penalized logistic regression \citep{LassoWu}, and PLasso \citep{pLasso}. To evaluate the performance of our method, In each simulation, we let $\mathcal{J}$ denote the set of true active variables and $\widehat{\mathcal{J}}_i$  the set of selected variables in the $i$-th replication, $i=1, 2, \ldots, 500$. The following metric is used to evaluate the performance of each method:
\begin{equation}\label{eqn.metric}
\varrho = \frac{1}{500}\sum_{i=1}^{500}\frac{|\widehat{\mathcal{J}}_i \cap \mathcal{J}|}{|\mathcal{J}|},
\end{equation}
which measures the proportion of active variables selected out of the total amount of true active variables. Such metric has been widely used in feature selection literature (see,  for instance, \cite{SIS} and \cite{pLasso}).

To validate the comparison, we use the Wilcoxon method to test whether the difference between DNN method and each of the other methods is statistically significant. As such, the following hypotheses are considered:
\begin{equation}\label{wilcoxon}
H_0: \varrho_{\textrm{DNN}}  = \varrho_{k} ~ \textrm{ versus } ~ H_a: \varrho_{\textrm{DNN}}  > \varrho_{k},
\end{equation}
where $\{\varrho_{1}, \varrho_2, \varrho_3, \varrho_4\} $ are the $\varrho$s associated with ATT, ISIS, LassoWu, andPLasso, respectively.  Denote by $p_k$ the p-value of the Wilcoxon test for a comparison between DNN and method $k\in \{1, 2, 3, 4\}$, and let
\begin{equation}\label{wilcoxon1}
p_{max} = \max_{k=1, 2, 3, 4}p_k.
\end{equation}
To test the statistical significance, we aim to show $p_{max} $ is less than $1\%$.

In each replication and for various combinations of $\rho, n$, and $p$, we utilize the five methods to conduct feature selection on the simulated data. The results, depicted in Table \ref{Tab:Sim1_classification}, showcase the average proportion of features selected by each method out of the 10 true active features over 500 replications, as defined in Equation \ref{eqn.metric}. The corresponding standard errors are presented in parentheses. Across different scenarios involving $\rho, n$, and $p$, distinctive patterns in performance emerge. Specifically, under the conditions of $\rho = 0, n = 200$, and $p = 1000$, the proposed DNN method exhibits a feature selection rate of 63\%, surpassing PLasso, which follows closely, while ATT achieves a rate of 38\%. As expected, an increase in the sample size positively influences the performance of all methods, with DNN leading the pack at 83\% feature selection when $n = 1000$. However, elevated dimensions in $p$ and/or heightened dependence with $\rho$ among the SNPs adversely affect the efficacy of feature selection methods. The p-values, consistently below 1\%, firmly establish that DNN consistently outperforms the other methods across all evaluated aspects.

\begin{table}[!t]
\centering
\caption{Results of Simulation 1: the averaged $\varrho$ over 500 replicates (with its standard error in parentheses) of various methods using different combinations of $\rho, n$, and $p$.  (+) means that $p_{\textrm{max}} < 0.01$, where $p_{\textrm{max}}$ is the maximum $p$-values for the six tests defined in \eqref{wilcoxon1}.}
\label{Tab:Sim1_classification}
{\begin{tabular}{@{}ccccccc@{}}
\hline
$\rho$ &($n, p$)     & ATT   & ISIS & LassoWu & PLasso & DNN \\
0 &(200, 1,000) &  0.38  & 0.45  & 0.51     & 0.55 & {\bf 0.63} (+) \\
  &            & (0.03)& (0.04)& (0.04)   &(0.05)& (0.04)  \\
0 &(500, 1,000) &   0.44  & 0.52  & 0.59     & 0.63 & {\bf 0.72} (+) \\
  &            & (0.04)& (0.04)& (0.05)   &(0.05)& (0.06)  \\
0 &(1,000, 1,000) & 0.52  & 0.60  & 0.66     & 0.71 & {\bf 0.83} (+) \\
  &            & (0.05)& (0.05)& (0.05)   &(0.06)& (0.06)  \\
\hline
\hline
$\rho$ &($n, p$)  & ATT   & ISIS & LassoWu & PLasso & DNN\\
0 &(200, 3,000) &  0.32  & 0.39  & 0.44     & 0.48 & {\bf 0.55} (+) \\
  &            & (0.03)& (0.04)& (0.03)   &(0.04)& (0.04)  \\
0 &(500, 3,000) &   0.39  & 0.45  & 0.50     & 0.55 & {\bf 0.65} (+) \\
  &            & (0.04)& (0.04)& (0.03)   &(0.04)& (0.04)  \\
0 &(1,000, 3,000) & 0.42  & 0.53  & 0.57     & 0.61 & {\bf 0.72} (+) \\
  &            & (0.04)& (0.05)& (0.05)   &(0.05)& (0.06)  \\
\hline
\hline
$\rho$ &($n, p$)   & ATT   & ISIS & LassoWu & PLasso & DNN \\
0.5 &(200, 1,000) &  0.33  & 0.41  & 0.45     & 0.50 & {\bf 0.59} (+) \\
  &            & (0.03)& (0.04)& (0.03)   &(0.04)& (0.04)  \\
0.5 &(500, 1,000) &   0.40  & 0.47  & 0.55     & 0.58 & {\bf 0.67} (+) \\
  &            & (0.04)& (0.04)& (0.04)   &(0.05)& (0.05)  \\
0.5 &(1,000, 1,000) & 0.48  & 0.55  & 0.60     & 0.66 & {\bf 0.74} (+) \\
  &            & (0.05)& (0.05)& (0.04)   &(0.05)& (0.05)  \\
\hline
\hline
$\rho$ &($n, p$)  & ATT   & ISIS & LassoWu & PLasso & DNN\\
0.5 &(200, 3,000) &  0.28  & 0.35  & 0.40     & 0.43 & {\bf 0.50} (+) \\
  &            & (0.03)& (0.03)& (0.03)   &(0.04)& (0.03)  \\
0.5 &(500, 3,000) &   0.35  & 0.44  & 0.46     & 0.51 & {\bf 0.60} (+) \\
  &            & (0.04)& (0.04)& (0.03)   &(0.04)& (0.04)  \\
0.5 &(1,000, 3,000) & 0.39  & 0.49  & 0.52     & 0.57 & {\bf 0.68} (+) \\
  &            & (0.04)& (0.05)& (0.04)   &(0.05)& (0.05)  \\
\hline
\hline
$\rho$ &($n, p$)   & ATT   & ISIS & LassoWu & PLasso & DNN \\
0.8 &(200, 1,000) &  0.30  & 0.37  & 0.42     & 0.47 & {\bf 0.55} (+) \\
  &            & (0.03)& (0.04)& (0.03)   &(0.04)& (0.04)  \\
0.8 &(500, 1,000) &   0.36  & 0.44  & 0.52     & 0.55 & {\bf 0.62} (+) \\
  &            & (0.04)& (0.04)& (0.04)   &(0.04)& (0.04)  \\
0.8 &(1,000, 1,000) & 0.43  & 0.51  & 0.56     & 0.61 & {\bf 0.70} (+) \\
  &            & (0.04)& (0.05)& (0.04)   &(0.05)& (0.04)  \\
\hline
\hline
$\rho$ &($n, p$)  & ATT   & ISIS & LassoWu & PLasso & DNN\\
0.8 &(200, 3,000) &  0.25  & 0.32  & 0.36     & 0.40 & {\bf 0.46} (+) \\
  &            & (0.03)& (0.03)& (0.03)   &(0.03)& (0.03)  \\
0.8 &(500, 3,000) &   0.32  & 0.41  & 0.41     & 0.47 & {\bf 0.55} (+) \\
  &            & (0.04)& (0.04)& (0.03)   &(0.04)& (0.04)  \\
0.8 &(1,000, 3,000) & 0.33  & 0.45  & 0.49     & 0.52 & {\bf 0.62} (+) \\
  &            & (0.04)& (0.05)& (0.04)   &(0.04)& (0.04)  \\
\hline
\end{tabular}}{}
\end{table}

\subsection{Experiment 2}
In this simulation, we utilize the \emph{Breast Cancer Coimbra Data Set} sourced from the UCI Machine Learning Repository (\href{https://archive.ics.uci.edu/ml/datasets/Breast+Cancer+Coimbra}{https://archive.ics.uci.edu/ml/datasets/Breast+Cancer+Coimbra}). The dataset comprises $9$ quantitative predictors and a binary response, denoting the presence or absence of breast cancer, with a sample size of $n=116$. While it is reasonable to assume a strong association between these $9$ predictors and the response, the specific impact on the response remains unknown. To evaluate our method, we adopt the following procedure: incorporating an additional $p-9$ irrelevant predictors from a standard normal distribution, independent of the response, with $p$ taking values of $1000, 5000, 10000$. Essentially, out of all $p$ predictors, only $9$ are active variables correlated with the response, rendering the rest irrelevant. The manner in which these $9$ variables influence the response remains unspecified. The results presented in Table \ref{Tab:unkonwn} encompass the statistic $\varrho$ based on 500 replications and its corresponding standard error. Additionally, we assess whether the maximum p-value for the six tests defined in \ref{wilcoxon1} is less than 1\% for $p = 1000, 5000$, and $10000$. An intriguing observation is that, irrespective of the dimensionality $p$, the DNN method consistently identifies all features. As the feature dimension $p$ expands, the selection precision of other methods diminishes. In contrast, our proposed method exhibits superior capability in selecting the correct active predictors, outperforming others across varied dimensionalities.

\begin{table}[!t]
\centering
\caption{Results of Simulation 3: the averaged $\varrho$ over 500 replicates (with its standard error in parentheses) of various methods using different $p$s.  (+) means that $p_{\textrm{max}} < 0.01$, where $p_{\textrm{max}}$ is the maximum $p$-values for the six tests defined in \eqref{wilcoxon1}. }
\label{Tab:unkonwn}
{\begin{tabular}{@{}cccccccc@{}}
\hline
($n, p$)      & ATT   & ISIS & LassoWu & PLasso & DNN \\
(116, 1,000)  &  0.17 & 0.20  & 0.19     & 0.18    & {\bf0.99} (+)\\
               & (0.02)& (0.02)& (0.2)   & (0.03)  &(0.2)\\
(116, 5,000)    & 0.16  & 0.18  & 0.16     & 0.15    &{\bf0.98}  (+)\\
              & (0.02)& (0.03)& (0.03)   & (0.03)  &(0.03)\\
(116, 10,000)    & 0.14  & 0.13  & 0.0.14     & 0.13    &{\bf0.98} (+)\\
             & (0.02)& (0.02)& (0.02)   & (0.02)  &(0.02)\\
\hline
\end{tabular}}{}
\end{table}
\subsection{Experiment 3}

\begin{figure}
\centering
\includegraphics[scale=.28]{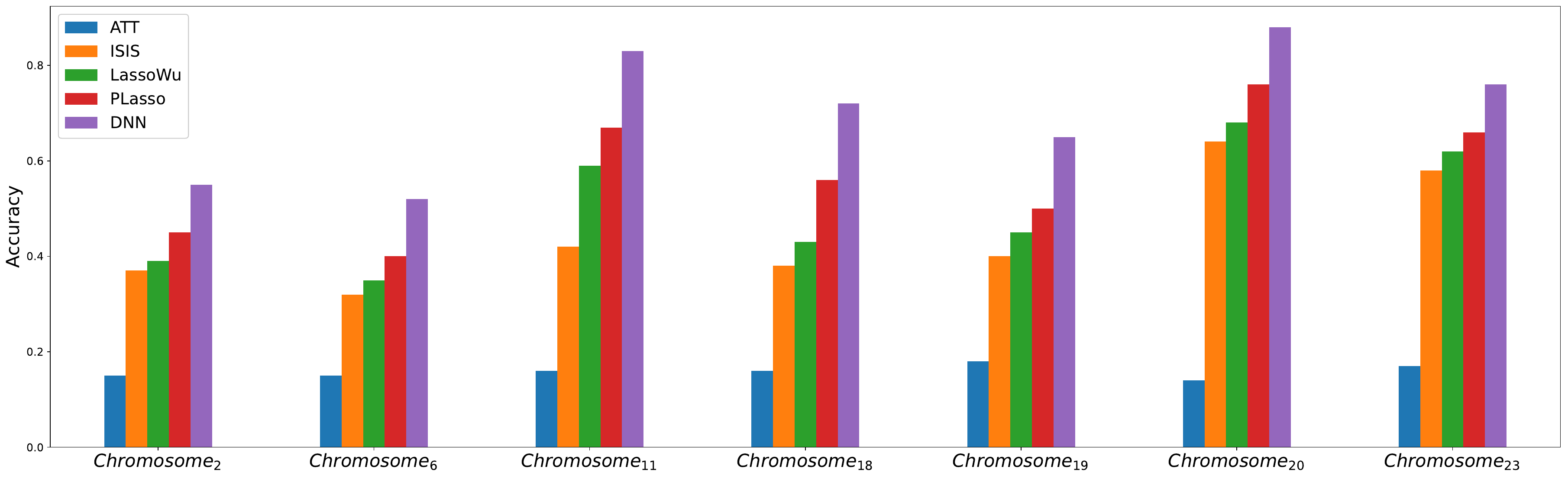}
\caption{The results of Experiment 3.}
\label{Sim3}
\end{figure}

In this simulation, we leveraged the robust capabilities of the GWAsimulator tool \citep{GWAsimulator}, a proficient C++ program meticulously designed to simulate genotype data originating from genomic SNP chips widely employed in GWAS. Renowned for its efficiency, GWAsimulator implements a rapid moving-window algorithm \citep{durrant2004linkage}, allowing for the simulation of comprehensive genome datasets catering to both case-control and population samples. Notably, this versatile program offers the added flexibility of simulating specific genomic regions as per user specifications, enhancing precision and customization in simulations. Specifically designed for case-control simulations, GWAsimulator empowers users to define various disease model parameters, including disease prevalence, the count of disease loci, and specific details for each locus such as location, risk allele, and genotypic relative risk. The program's adaptability extends further to enable users to focus simulations on particular genomic regions, providing a tailored and nuanced approach. In the context of our simulation, GWAsimulator played a pivotal role in generating the GWAS data. A comprehensive summary of the simulated data is presented in Table 3. Subsequently, we applied five distinct approaches to analyze this dataset, and the outcomes are succinctly encapsulated in Figure \ref{Sim3}. Notably, the results underscore the superior selection accuracy achieved by the DNN approach in comparison to the other methods.

\begin{table}[]
\caption{The disease locus position is the position in the input phased file, not the physical position on a
chromosome. For example, the disease locus in chromosome 19 is the 2885th SNP, with allele 1 as the disease-risk allele. The first relative risk, $RR_1$, is
the risk ratio of the genotype with one copy of the risk allele versus that with zero copy of the risk
allele. Similarly, the second relative risk, $RR_2$, is the risk ratio of the genotype with two copies of the
risk allele versus that with zero copy of the risk allele. If it is “M”, then the multiplicative effect is
assumed and $RR_2 = RR_1^2$. If it is “D”, then the dominance effect is assumed, and $RR_2 = RR_1$. For the recessive effect, specify 1.0 for $RR_1$. All relative risks should be $\geq1$.}
\centering
\begin{tabular}{|c|c|c|c|c|c|c|c|}
\hline
\thead{Chromosome\\Number} & \thead{Position} & \thead{Disease\\Variant Allele} & \thead{The First\\Genotypic Relative \\Risks} & \thead{The Second\\Genotypic Relative \\Risks} & \thead{Start\\Position} & \thead{End\\Position} &\thead{Dimensionality} \\ \hline
2  & 10714 & 0 & 1.1 & D   & 10000 & 12000 & 25215\\ \hline
6  & 4322  & 1 & 1.0 & 1.1 & 3000  & 5600  & 20269\\ \hline
11 & 9067  & 1 & 1.5 & M   & 8000  & 10000 & 14520\\ \hline
18 & 9659  & 1 & 1.1 & M   & 6000  & 10000 & 10441\\ \hline
19 & 2885  & 1 & 1.5 & M   & 1000  & 4000  & 5789\\ \hline
20 & 3357  & 0 & 1.1 & 2.0 & 1000  & 5000  & 7802\\ \hline
23 & 7607  & 0 & 1.5 & 1.5 & 7000  & 9000  & 9120\\ \hline
\end{tabular}
\end{table}

\section{Conclusion}\label{Secconclusion}
In summary, our extended deep neural network approach emerges as a robust and adaptive solution for ultra-high-dimensional feature selection within GWAS data. By refining Mirzaei et al.'s (2020) method, we have effectively addressed the unique challenges associated with ultra-high-dimensional and small-sample setups prevalent in genomics research. The integration of a Frobenius norm penalty into the student network serves as a pioneering enhancement, significantly broadening the method's applicability to GWAS datasets characterized by intricate structures and limited sample sizes. Beyond its technical advancements, our approach stands out for its flexibility, demonstrated through comprehensive experiments across diverse genomic scenarios. The method's proficiency in unraveling complex relationships embedded in GWAS datasets positions it as a promising solution for gaining valuable insights into genetic associations with diseases. This work makes a substantial contribution to advancing feature selection methodologies, explicitly tailored to the specific demands of ultra-high-dimensional genomics research. As big genomics data continues to evolve, our extended deep neural network approach emerges as a potent tool for researchers seeking accurate and interpretable feature selection in the complex landscape of GWAS.



\clearpage
\bibliographystyle{apalike}
\bibliography{references}  






\end{document}